\documentclass[untypical,portrait]{baposter}

\usepackage{relsize}	
\usepackage{url}	
\usepackage{graphicx} 
\usepackage{epstopdf}	

\graphicspath{{pix/}}	
\definecolor{bordercol}{RGB}{240,240,120}
\definecolor{headercol1}{RGB}{256,256,256}
\definecolor{headercol2}{RGB}{256,256,180}
\definecolor{headerfontcol}{RGB}{0,0,0}
\definecolor{boxcolor}{RGB}{2256,256,256}

\begin{document}
\typeout{Poster rendering started}

\background{
	\begin{tikzpicture}[remember picture,overlay]%
	\draw (current page.north west)+(-2em,2em) node[anchor=north west]
	{\includegraphics[height=1.1\textheight]{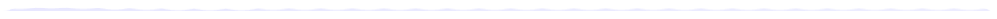}};
	\end{tikzpicture}
}

\begin{poster}{
	grid=false,
	borderColor=bordercol,
	headerColorOne=headercol1,
	headerColorTwo=headercol2,
	headerFontColor=headerfontcol,
	boxColorOne=boxcolor,
	headershape=roundedright,
	headerfont=\Large\sf\bf,
	textborder=rectangle,
	background=user,
	headerborder=open,
  boxshade=plain
}
{
	Eye Catcher, empty if option eyecatcher=false - unused
}
{\sf\bf
	Chemical abundance analysis of symbiotic giants.\\ 
\vspace{0.2em}
        RW Hya, SY Mus, BX Mon, and AE Ara
}
{
	\vspace{1em} Cezary Galan$^1$, Joanna Mikolajewska$^1$, Kenneth H. Hinkle$^2$, Miroslaw R. Schmidt$^3$, and Mariusz Gromadzki$^4$\\
\vspace{0.5em}
{\smaller $^{1}$N. Copernicus Astronomical Center, Bartycka 18, PL-00-716 Warsaw, Poland, (E-mail: cgalan@camk.edu.pl)\\
$^{2}$National Optical Astronomy Observatory, P.O. Box 26732, Tucson, AZ 85726, USA\\
$^{3}$N. Copernicus Astronomical Center, Rabia\'nska 8, PL-87-100 Toru\'n, Poland\\
$^{4}$Departamento de F{\'i}sica y Astronom{\'i}a, Universidad de Valpara{\'i}so, Av. Gran Breta\~{n}a 1111, Playa Ancha, Casilla 5030, Chile}
}

\headerbox{Introduction}{name=intro,column=0,row=0}{
Symbiotic stars are long period binary systems, composed of two evolved and
strongly interacting stars: a red giant and a hot and luminous, typically
white dwarf, companion surrounded by an ionized nebula.  Mass loss from the
giant undergoes accretion to the compact object via wind and$/$or Roche lobe
overflow (e.g., \cite{Pod2007}, \cite{Mik2012}) resulting in the formation
of accretion discs and jets (e.g., \cite{Sol1985}, \cite{Tom2003},
\cite{Ang2011}).  The hot companion was previously a red giant and this mass
transfer episode, in the opposite direction, should have left traces in the
chemical composition of the current red giant.  Symbiotic stars can be
progenitors for Supernovae Type Ia and can be responsible for several or
perhaps up to about thirty percent of these events (e.g., \cite{Dil2012},
\cite{Mik2012}).\\
The complex structure including the many kinds of interactions make
symbiotic stars excellent laboratories for studying various aspects of the
late stages of binary evolution including those that impact the chemical
evolution of the Galaxy and the formation of stellar populations.  Knowledge
of the chemical composition of symbiotic giants is essential to advancing
our understanding but unfortunately reliable determinations exist in a few
cases.  Thus far, analysis of the photospheric chemical abundances have been
performed for four only normal S-type symbiotic giants (V2116\,Oph
\cite{Hin2006}, T\,CrB, RS\,Oph \cite{Wal2008} and CH\,Cyg \cite{Sch2006}). 
In all these cases solar or nearly solar metallicities were found.  Some
information about chemical composition is also available for roughly a dozen
so called yellow symbiotic systems.  These are S-type symbiotics containing
a giant of K or G spectral type that is metal poor with overabundant
s-process elements (AG\,Dra \cite{Smi1996}, BD-21\,3873, \cite{Smi1997},
Hen\,2-467 \cite{Per1998}, CD-43\,14304, Hen\,3-1213, Hen\,3-863, StHA\,176
\cite{Per2009}).  Compositions also have been measured for members of the
small D' subclass of symbiotic stars that contain fast rotating G-type
giants and warm dust shells.  These have solar metallicities with
overabundant s-process elements (StHA\,190 \cite{Smi2001}, HD\,330036,
AS\,201 \cite{Per2005}).  The number of objects is too small for statistical
considerations.\\
We have started a chemical composition analysis for a sample of over 30
symbiotic systems.  Here we present the results obtained thus far for four
objects: RW\,Hya, SY\,Mus, BX\,Mon and AE\,Ara.
}

\headerbox{Observations and data reduction}{name=data,column=0,below=intro}{

The observational data are high-resolution ($R = \lambda/\Delta\lambda \sim 50000$; S/N
$\geq$\,100), near-IR spectra collected with the Phoenix cryogenic echelle
spectrometer on the 8\,m Gemini South telescope during the years 2003--2006. 
The spectra cover narrow wavelength ranges ($\sim$100\AA) located in the $H$ and
$K$ photometric bands at mean wavelengths 1.563\,$\mu$m, 2.225\,$\mu$m, and
2.361\,$\mu$m (hereafter $H$, $K$, and $K'$-band spectra, respectively). 
The $H$-band spectra contain molecular CO and OH lines, $K$-band spectra CN
lines, with both ranges useful for abundances of carbon,
nitrogen and oxygen and elements around iron peak: Sc, Ti, Fe, Ni.  The
$K'$-band spectra are dominated by strong CO features that enable measurement
of the $^{12}$C$/^{13}$C isotopic ratio.  The spectra were extracted and
wavelength calibrated using standard reduction techniques \cite{Joy1992} and
all were heliocentric corrected.  In all cases telluric lines were either
not present in the interval observed or were removed by reference to a hot
standard star.  The Gaussian instrumental profile is in all cases about 
6\,km\,s$^{-1}$ FWHM corresponding to instrumental
profiles of 0.31\AA\ , 0.44\AA\, and 0.47\AA\ in the case of the $H$-band,
$K$-band, and $K'$-band spectra, respectively.  The log of our spectroscopic
observations is given in the table below.

\vspace{-0.4em} 
\begin{center}
\includegraphics[width=0.85\linewidth]{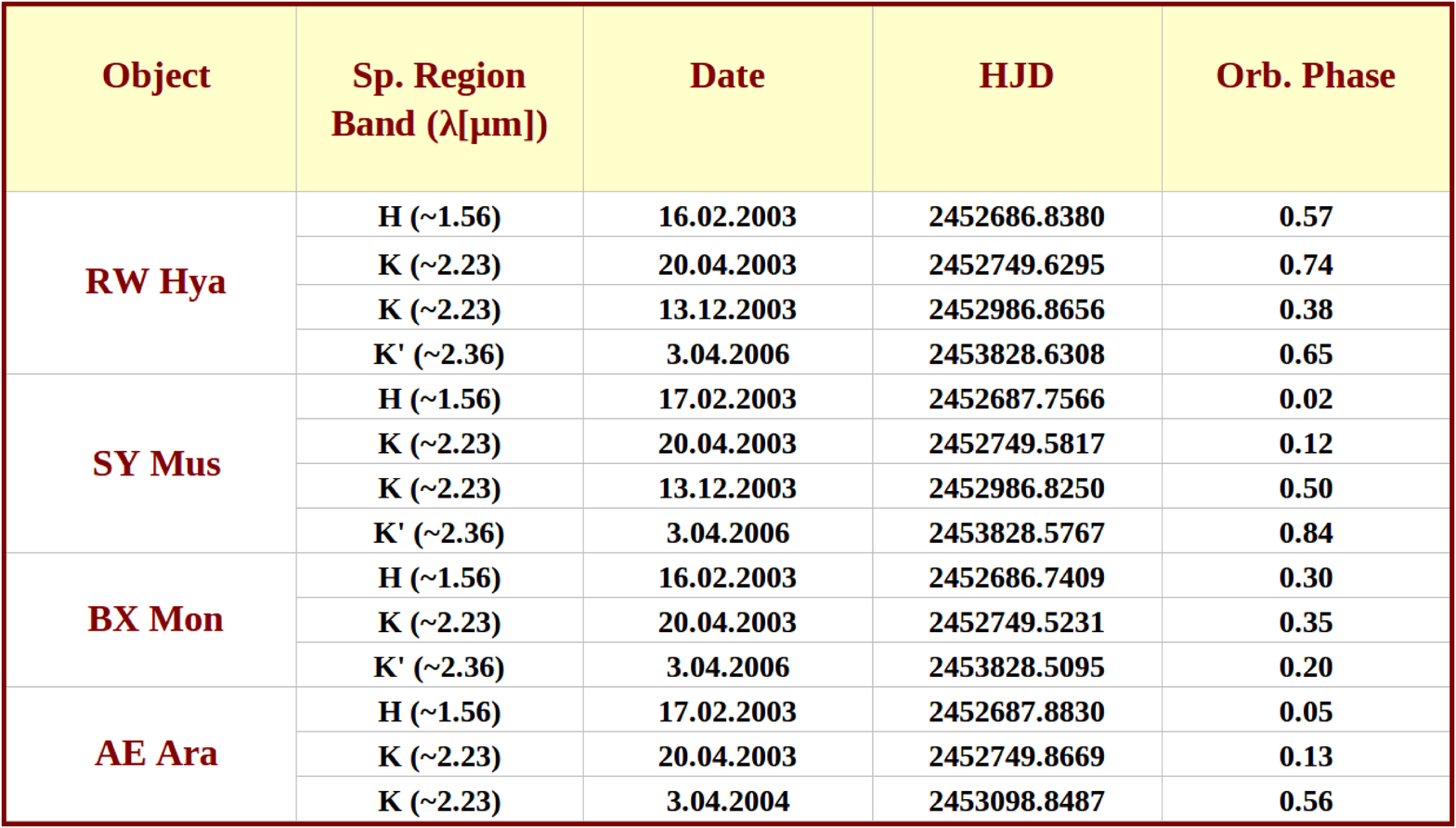}
\end{center}
}

\headerbox{References}{name=references,column=0,below=data}{
\smaller													
\vspace{-0.4em} 										
\bibliographystyle{plain}							
\renewcommand{\section}[2]{\vskip 0.05em}		

}

\headerbox{Methods}{name=methods,span=1,column=1}{

\begin{center}
\bf{The standard LTE analysis}
\end{center}
\vspace{-0.9em}
The abundance analyses were performed by fitting synthetic spectra to
observed ones with a method similar to that used for CH Cyg by Schmidt et
al.\,\cite{Sch2006}.  Standard \emph{LTE} analysis and \emph{MARCS} model
atmospheres developed by Gustafsson et al.\,\cite{Gus2008} were used for the
spectral synthesis.  The code \emph{WIDMO} developed by M.R. Schmidt was
used to calculate synthetic spectra over the observed spectral regions.  To
perform a $\chi^2$ minimization, a special overlay was developed on the
\emph{WIDMO} code with use of the simplex algorithm \cite{Bra1998}.  Use of
this procedure, in our case, enables an improvement of about ten times in
the computation efficiency.  The atomic data were taken from the \emph{VALD}
database \cite{Kup1999} in the case of $K$- and $K'$-band regions and from
the list given by Melendez \& Barbuy \cite{Mel1999} for the $H$-band region. 
For the molecular lines we used the line lists published by Goorvitch
\cite{Goo1994} (CO) and from CD-ROMs of Kurucz \cite{Kurucz} (CN and OH).

\vspace{-0.5em}
\begin{center}
\bf{Input stellar parameters}
\end{center}
\vspace{-0.9em}

The input effective temperatures $T_{\rm{eff}}$ were estimated from the
known spectral types \cite{Mur1999} adopting the calibration by Richichi et
al.\,\cite{Ric1999} and van Belle et al.\,\cite{Bel1999} (see table below). 
From the infrared colors and color excesses we obtained intrinsic colors. 
Using the Kucinskas et al.\,\cite{Kuc2005}
$T_{\rm{eff}}$--$\log{g}$--$color$ relation for late-type giants, we
estimated surface gravities and effective temperatures that are in good
agreement with those from the spectral types.  We then used model
atmospheres for the estimated values of surface gravities and the effective
temperature set to $T_{\rm{eff}}$=3700\,K for RW\,Hya and
$T_{\rm{eff}}$=3400\,K for SY\,Mus, BX\,Mon and AE\,Ara.  The
macroturbulence velocity $\zeta_{\rm{t}}$ was set at 3\,km$/$s, a value
typical for the cool red giants (e.g.  \cite{Fek2003}).

\vspace{-0.4em}
\begin{center}
\includegraphics[width=\linewidth]{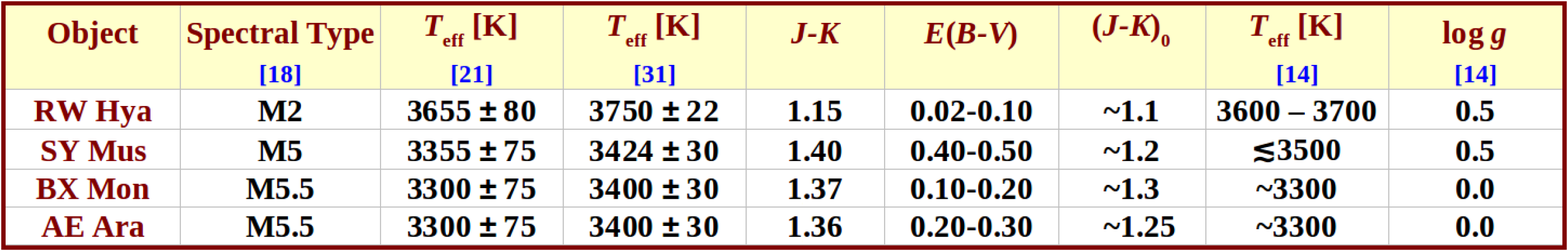}
\end{center}
\vspace{-0.9em}

The radial velocities for individual spectra were obtained with a
cross-correlation technique similar to that of Carlberg et
al.\,\cite{Car2011} but with synthetic spectra as the template.  Rotational
velocities were estimated via direct measurement of the FWHM of the 6
relatively strong, unblended atomic lines (Ti\,I, Fe\,I, Sc\,I) present in
the $K$-band region, the same lines used by Fekel et al.\,\cite{Fek2003}. 
The resulting values of the radial and rotational velocities were set as a
fixed parameters in our solutions.

\vspace{-0.5em}
\begin{center}
\bf{Procedure to derive abundances}
\end{center}
\vspace{-0.9em}

The following procedure was adopted to carry out the abundance solution.
\begin{itemize}
\vspace{-0.4em}
\item Estimation of the initial values of the abundance parameters: 
\vspace{-0.4em}
\item [--] initially the solar composition \cite{Asp2009} was adopted
\vspace{-0.4em}
\item [--] fitting by eye, alternately the OH, CO, CN and atomic lines
\vspace{-0.4em}
\item Building of the $n+1$, $n$ dimensional sets of free parameters -- the
so called simplex
\vspace{-0.4em}
\item Minimization with the simplex algorithm:
\vspace{-0.4em}
\item [--] 9 different simplexes were used, with different microturbulent
velocity $\xi_{\rm{t}}$ values (sampled in the range 1.2--2.6 km$/$s) to
obtain optimal fit to $H$- and $K$-band spectra
\vspace{-0.4em}
\item [--] searching for $^{12}$C$/^{13}$C by fitting to $K'$-band spectrum
\vspace{-0.4em}
\item [--] reconciliation of $^{12}$C and $^{12}$C$/^{13}$C within $\leq$3
iterations
\end{itemize}
\vspace{-0.4em}

}

\headerbox{Results}{name=results,span=1,column=1,below=methods}{

The final derived abundances for CNO elements and atomic lines (Sc\,I,
Ti\,I, Fe\,I, Ni\,I) on the scale $\log{\epsilon}(X) = log{N(X) N(H)^{-1}} +
12.0$, the isotopic ratios $^{12}$C/$^{13}$C, microturbulences
$\xi_{\rm{t}}$ and projected rotational velocities $V_{\rm{rot}} \sin{i}$,
are summarized in the table below together with formal uncertainties.  Our
analysis of the chemical abundances reveals a significantly subsolar
metallicity (Me$/$H $\sim -0.75$) for RW\,Hya, slightly subsolar
metallicities in BX\,Mon and AE\,Ara, and an approximately solar metallicity
in SY\,Mus.  The $^{12}$C$/^{13}$C isotopic ratios are low: $\sim$6,
$\sim$10, and $\sim$9, for RW\,Hya, SY\,Mus, and BX\,Mon respectively.

\vspace{-0.9em}
\begin{center}
\includegraphics[width=\linewidth]{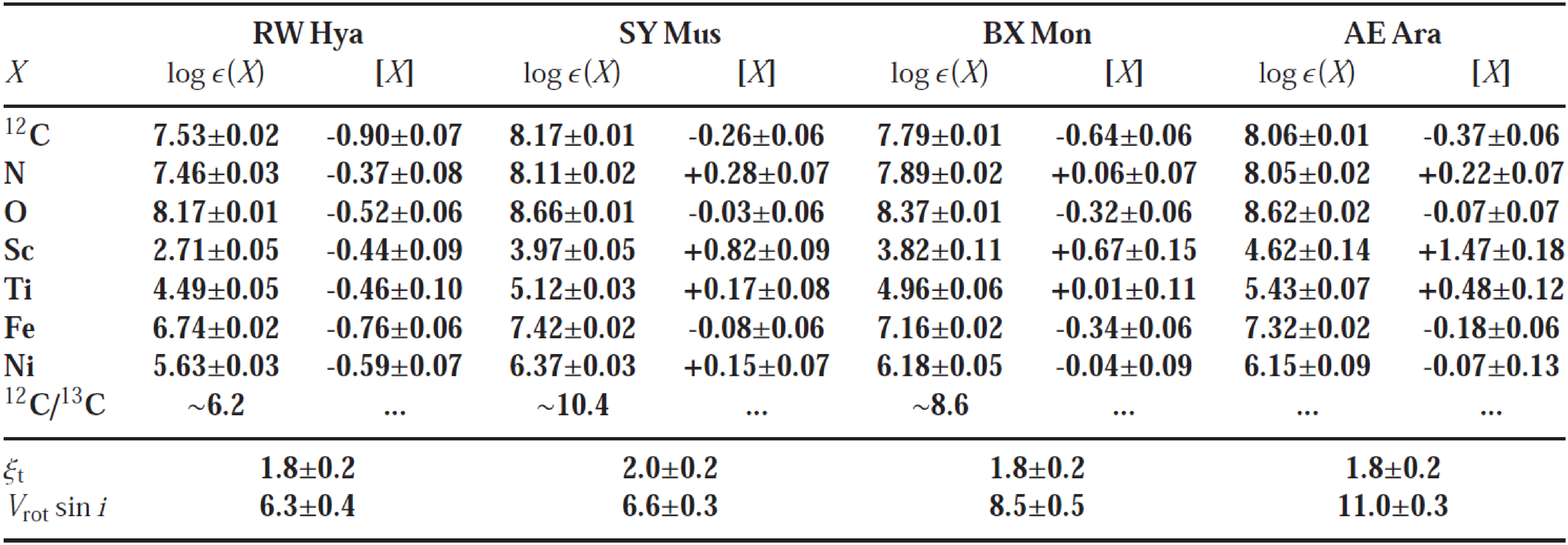}
\end{center}
\vspace{-0.75em}

The final synthetic fits to the spectra of RW\,Hya and BX\,Mon:
\vspace{-0.4em}
\begin{center}
	\includegraphics[width=0.495\linewidth]{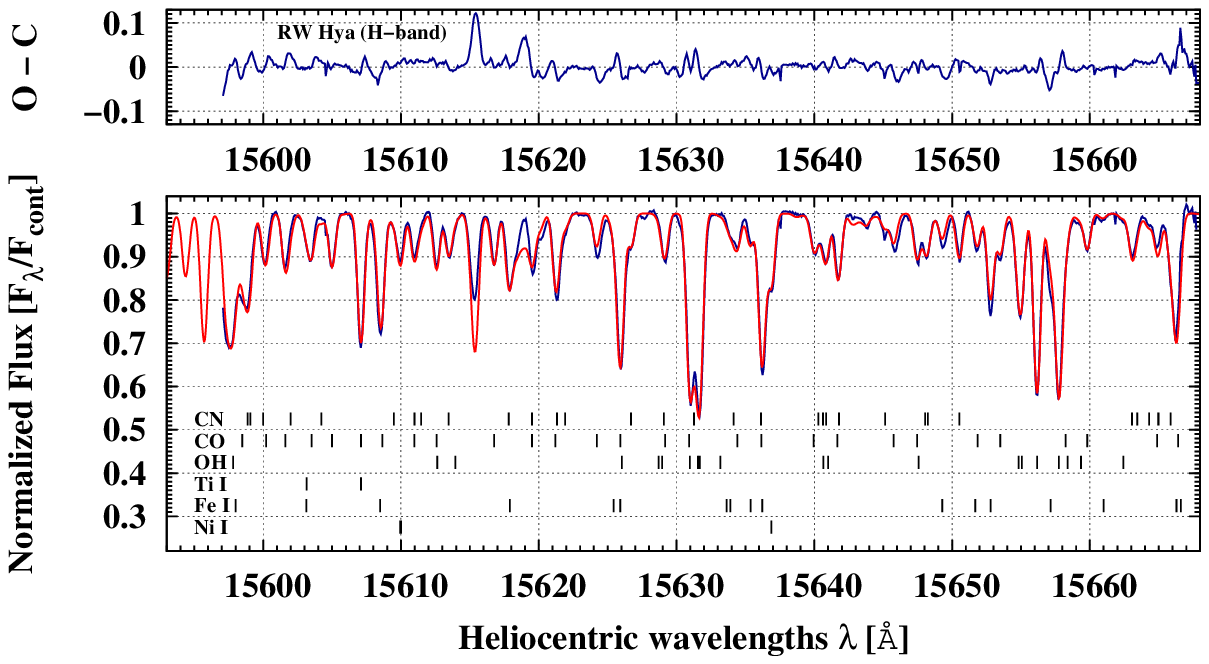}
	\includegraphics[width=0.495\linewidth]{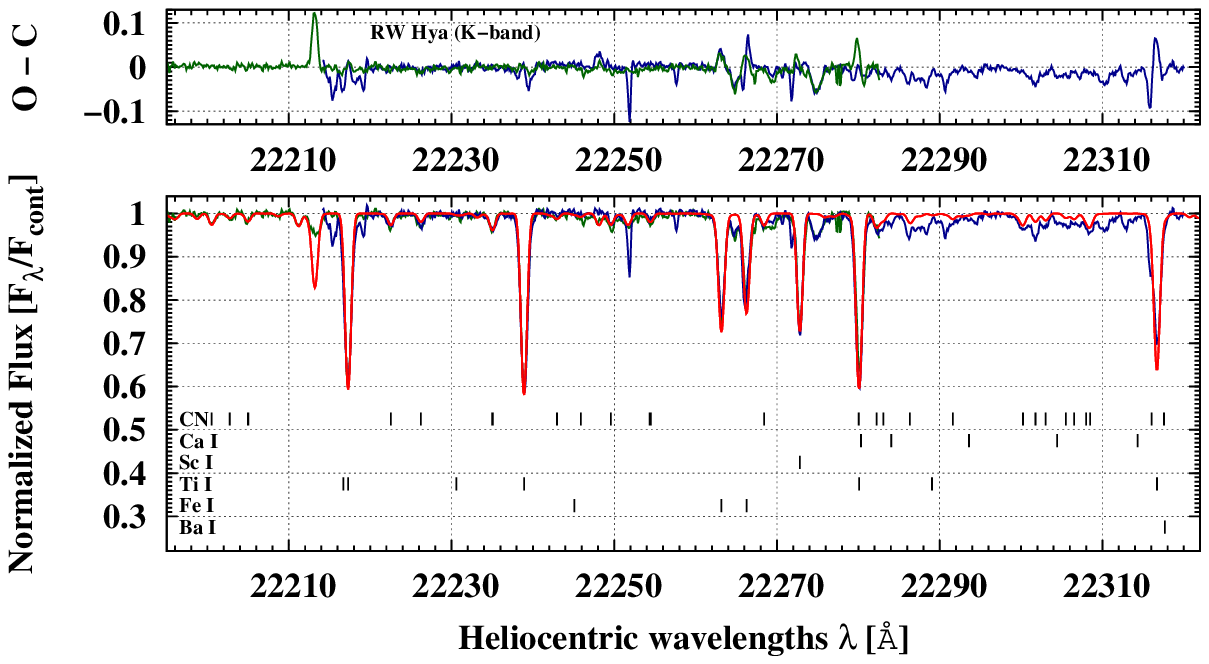}
\end{center}

\vspace{-1.5em}
\begin{center}
	\includegraphics[width=0.495\linewidth]{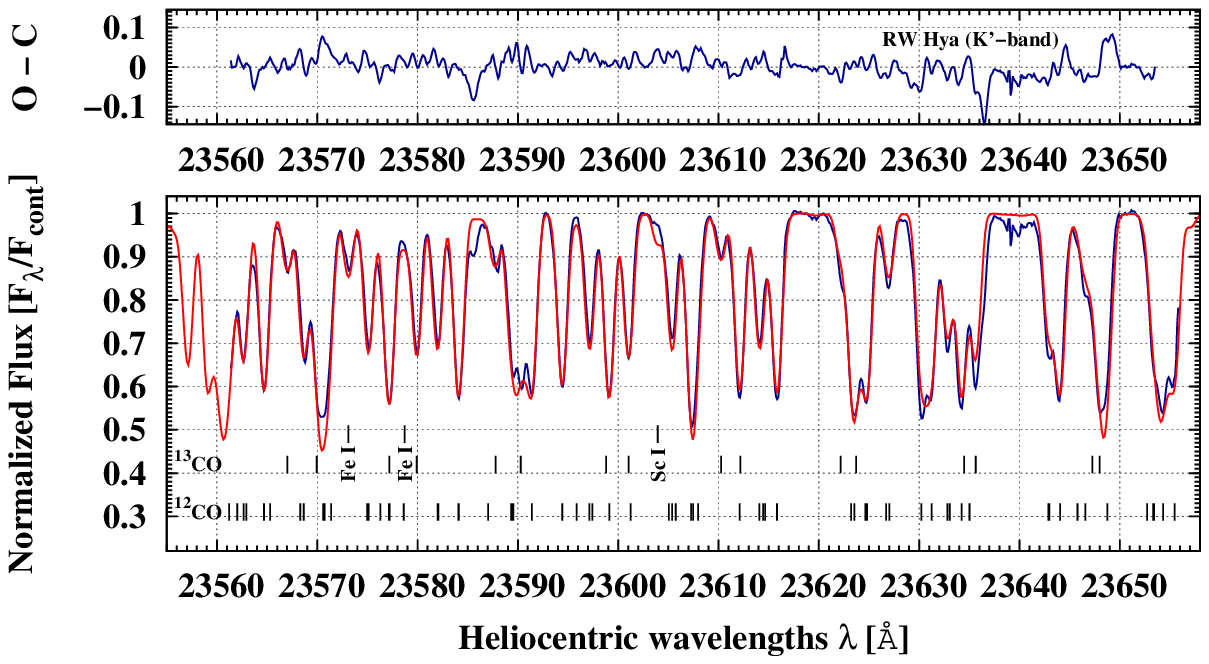}
	\includegraphics[width=0.495\linewidth]{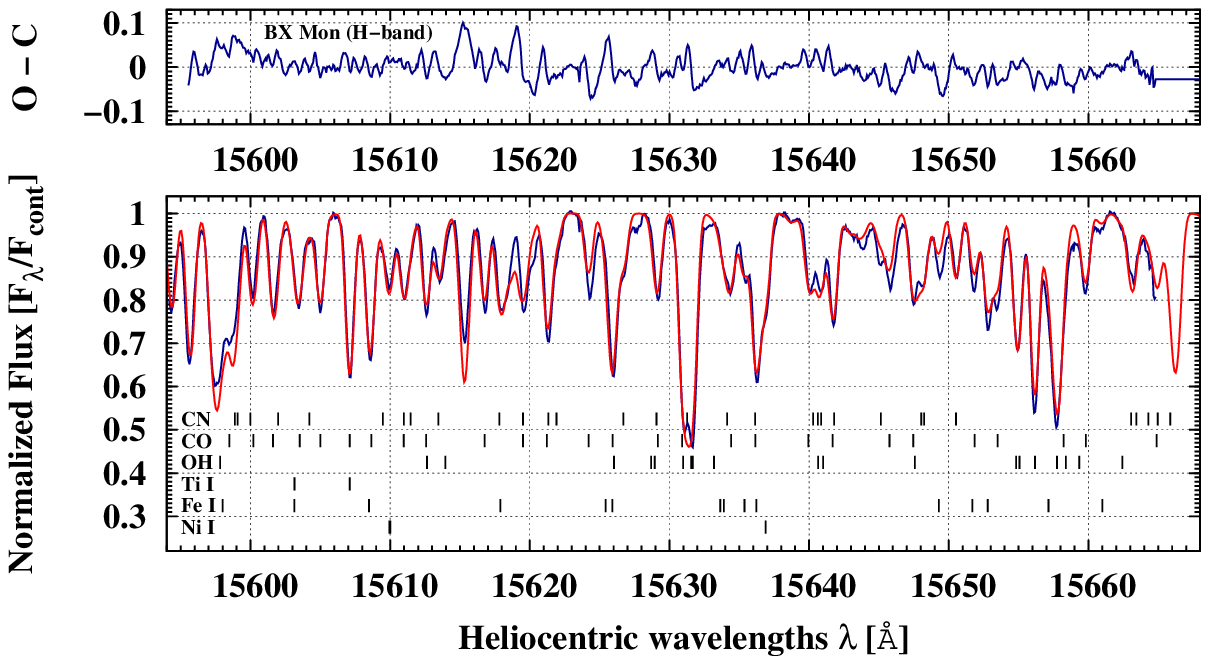}
\end{center}

\vspace{-1.7em}
\begin{center}
	\includegraphics[width=0.495\linewidth]{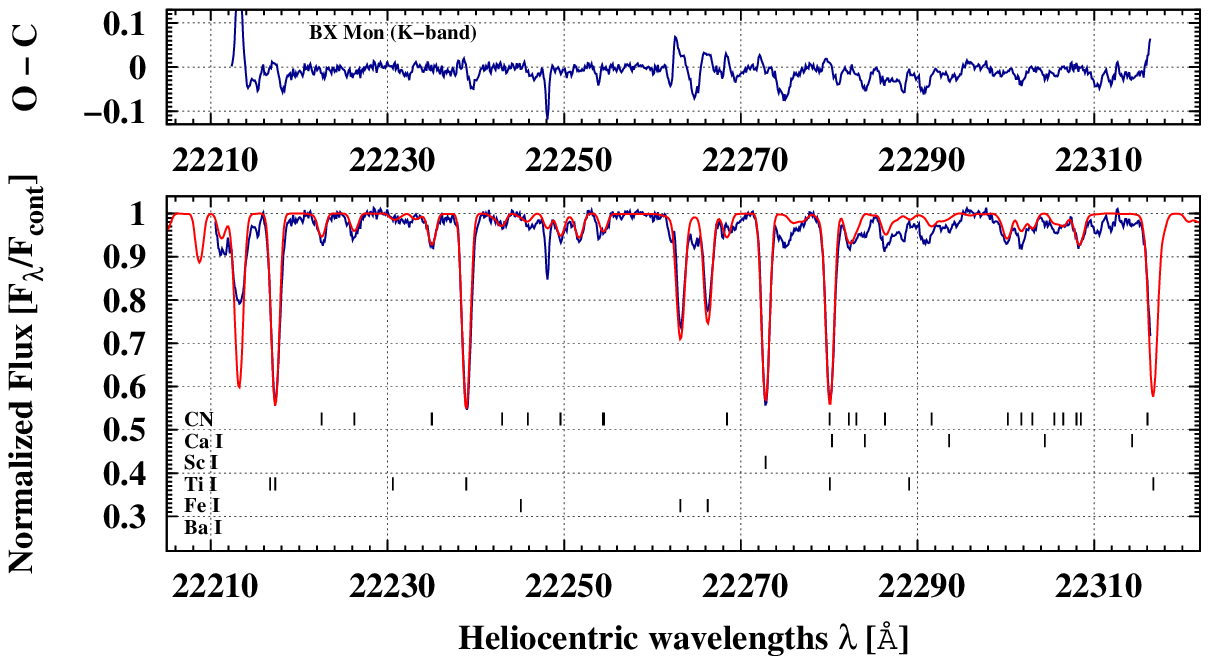}
	\includegraphics[width=0.495\linewidth]{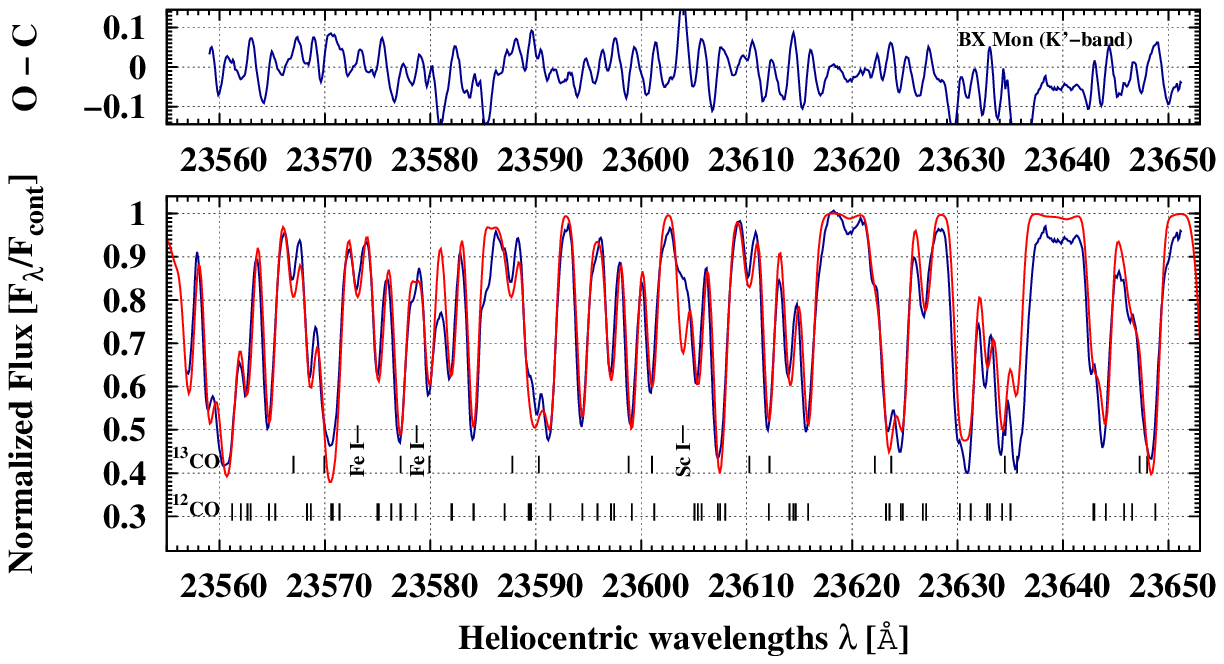}
\end{center}

}

\headerbox{Symbiotic stars in Galactic populations}{name=populations,span=1,column=2}{

The chemical composition analysis of the four symbiotic giants enables a
comparison of their chemical properties to those of Galactic stellar
populations previously obtained by various authors.  In the table below are
abundances and estimates of their ratios obtained by us for RW\,Hya,
SY\,Mus, BX\,Mon and AE\,Ara and those obtained by Schmidt et
al.\,\cite{Sch2006} for CH\,Cyg.

\vspace{-0.5em}
\begin{center}
\includegraphics[width=\linewidth]{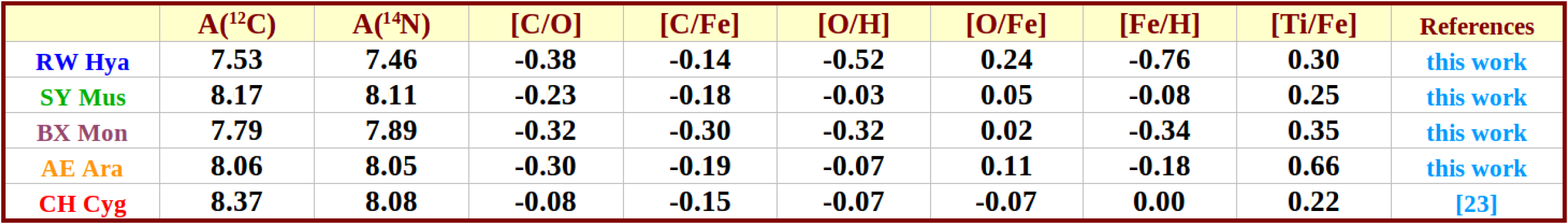}
\end{center}
\vspace{-0.9em}

We show the positions of these stars in the diagrams below with large dots
in blue, green, purple, orange, and red for RW\,Hya, SY\,Mus, BX\,Mon,
AE\,Ara, and CH\,Cyg respectively (according to the convention in the table
above).\\

\vspace{-0.6em}
Bensby \& Feltzing (fig.\,11--\cite{Ben2006}) present an analysis of the
trends of [C$/$Fe] and [C$/$O] versus [Fe$/$H] and [O$/$H] present in
various Galactic populations as a result of chemical evolution.  While the
[C$/$Fe]--[Fe$/$H] relation is totally merged and flat the others show clear
separation between thin- (open circles) and thick-disc (filled circles and
triangles) populations.  The uncertainties in the positions are quite large
($\sim$ 0.2-0.3), nevertheless we can see the RW Hya system (proper motions
suggest membership of the thick-disc or halo populations), placed far away
from the positions of the other systems, that in most cases seem fall into
regions dominated by thin-disc stars.

\vspace{-0.8em}
\begin{center}
\includegraphics[width=\linewidth]{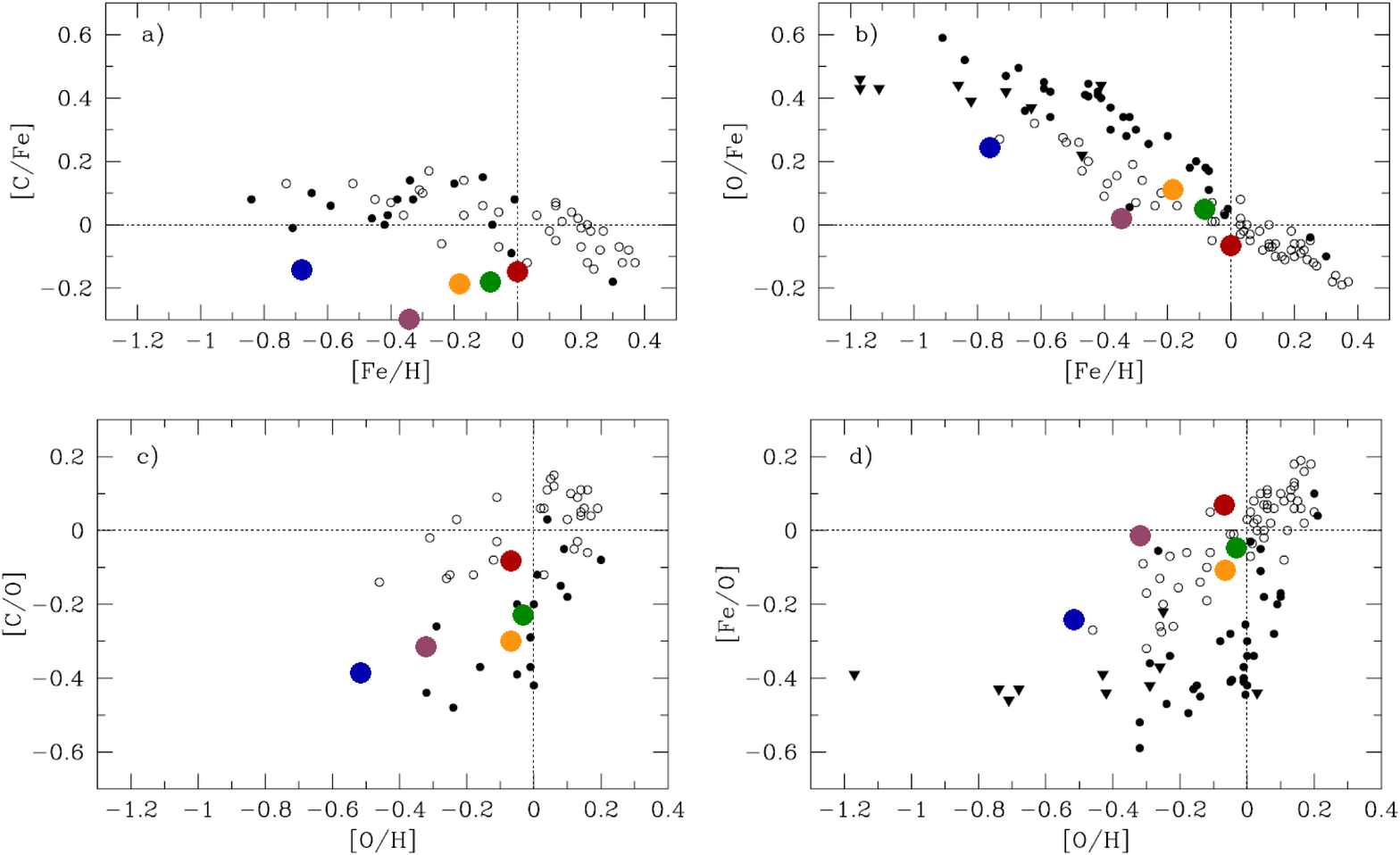}
\end{center}
\vspace{-0.9em}

On the left below is shown the relation [Ti$/$Fe]--[Fe$/$H] from Cunha \&
Smith (2006, fig.\,11--\cite{Cun2006}) for stars of different populations:
disc stars (open, blue circles), LMC (magenta squares), Sculptor (magenta
circles), Galactic bulge (pentagons with errorbars).  The position of
AE\,Ara on this diagram appears to confirm membership in the bulge
population.  On the right the CN cycle is shown (fig.\,4--\cite{Cun2006}). 
All of our stars analyzed so far fall into the $^{14}$N-rich zone.  This is
further evidence, in addition to the low $^{12}$C$/^{13}$C isotopic ratios,
that the first-dredge up has occurred in these objects.

\vspace{-0.7em}
\begin{center}
\includegraphics[width=\linewidth]{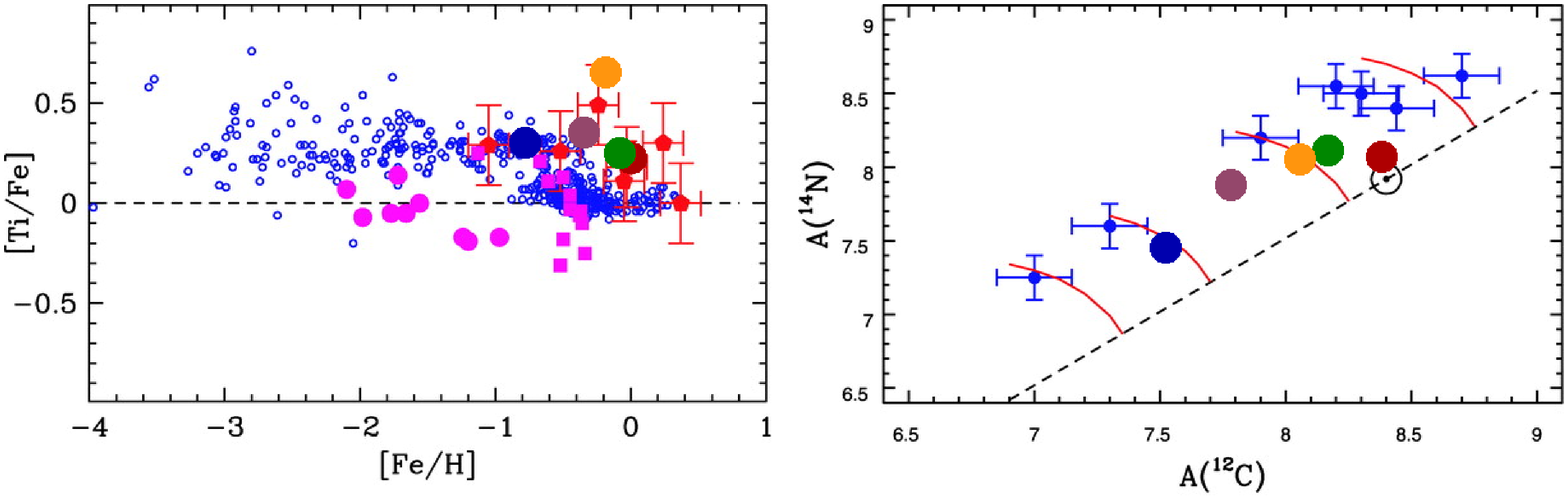}
\end{center}
\vspace{-0.9em}

The Galactic coordinates, distances and proper motions for the four
symbiotic systems enable us to estimate their Galactic velocities (U,V,W -
table below) and to analyze their position on the Toomre diagram (figure
below).  We used fig.\,1 from Feltzing et al.\,\cite{Fel2003} with denoted
positions of thin- (open circles), thick-disc (filled circles and open
squares), and halo stars (asterisks).  The RW\,Hya system is placed in the
extended thick-disk contrary to the other three symbiotic systems that are
placed in the thin-disc population.

\vspace{-0.5em}
\begin{center}
\includegraphics[width=\linewidth]{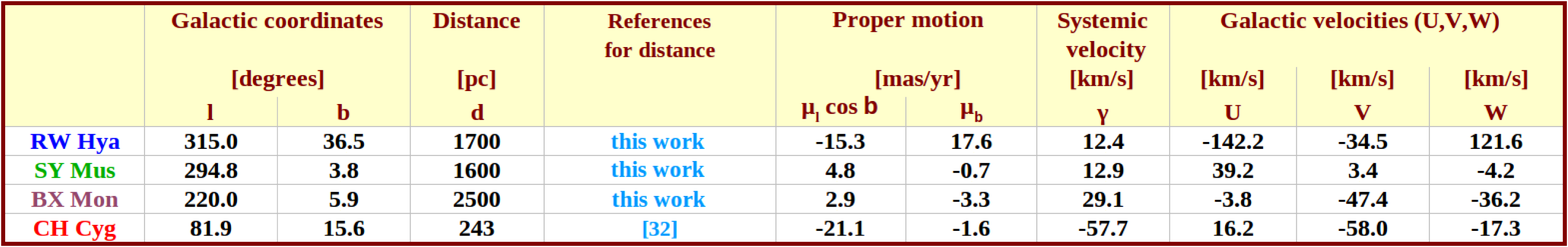}
\end{center}
\vspace{-1.2em}
\begin{center}
\includegraphics[width=0.84\linewidth]{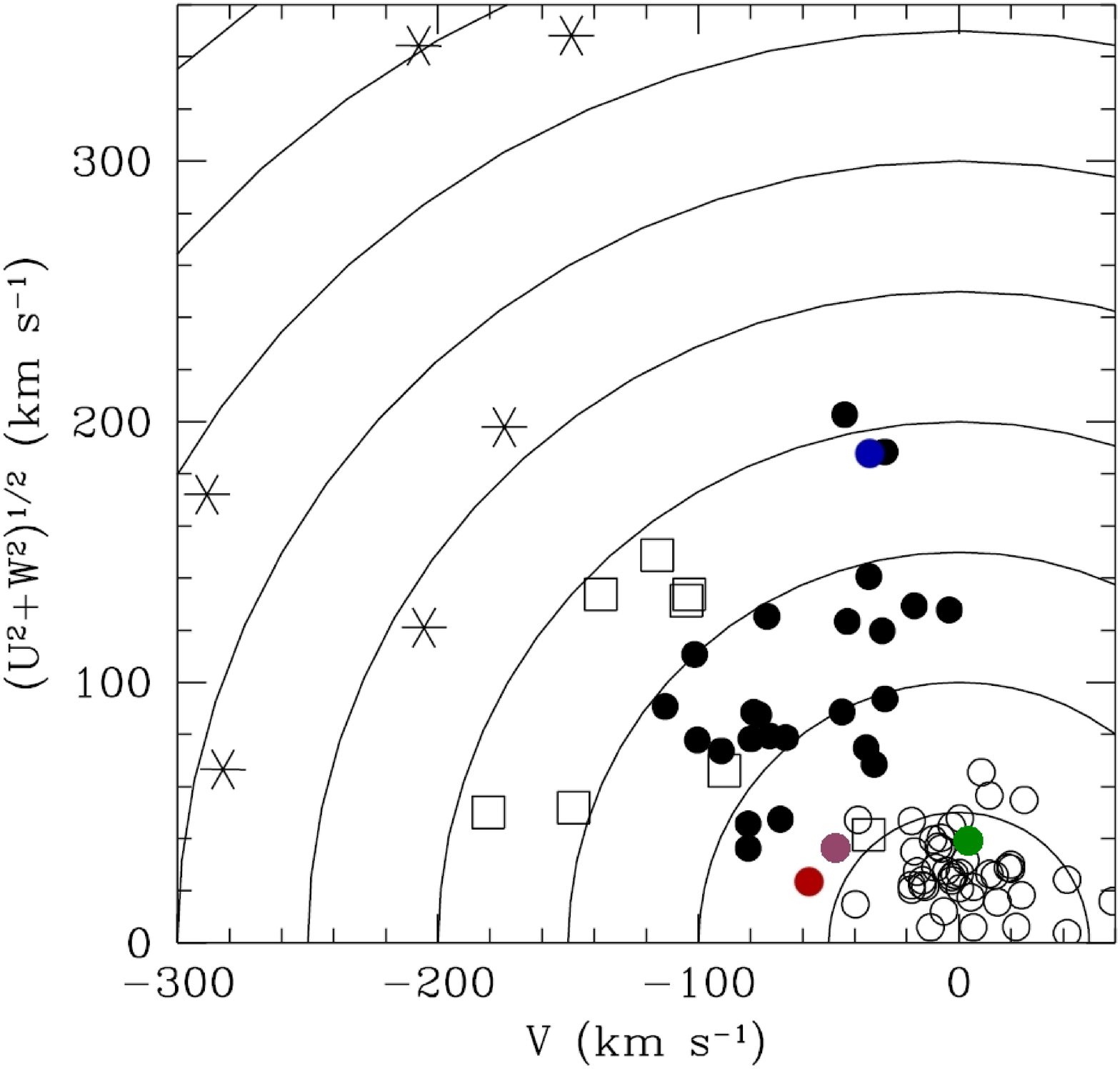}
\end{center}
\vspace{-0.8em}

We expect that an analysis like that presented above when applied to a
statistically significant sample will allow a better understanding of the
chemical evolution of symbiotic systems in various stellar populations.
}

\headerbox{Acknowledgements}{name=acknowledgements,column=2,below=populations}{
\smaller						
\vspace{-0.4em}			
This study is partly supported by the Polish National Science Centre grants  
No DEC-2011/01/B/ST9/06145 and No DEC-2013/08/S/ST9/00581.  MG is financed 
by the GEMINI-CONICYT Fund allocated to the project 32110014.  CG gratefully
acknowledges A.  Smith for careful reading of the manuscript.
}

\end{poster}
\end{document}